\documentclass[aps,prx,reprint,nofootinbib,preprintnumbers,amsmath,amssymb,amsfonts]{revtex4-2}

\usepackage[utf8]{inputenc}
\usepackage{graphicx}
\usepackage{hyperref}

\newcommand{\cdstrong}{c_d^{\textrm{strong}}}
\newcommand{\cdTCC}{c_d^{\textrm{TCC}}}

\begin{document}

\title{Asymptotic Observables and the Swampland}

\author{Tom Rudelius}
\email{rudelius@berkeley.edu}
\affiliation{Physics Department, University of California, Berkeley CA 94720 USA}

\date{\today}

\begin{abstract}
We show that constraints on scalar field potentials and towers of light massive states in asymptotic limits of scalar field space (as posited by the de Sitter Conjecture and the Swampland Distance Conjecture, respectively) are correlated with the prospects for defining asymptotic observables in expanding FRW cosmologies. The observations of a ``census taker'' in an eternally inflating cosmology are further related to the question of whether certain domain walls satisfy a version of the Weak Gravity Conjecture. This suggests that answers to fundamental questions about asymptotic observables in cosmology could help shed light on the Swampland program, and vice versa.
\end{abstract}


\maketitle

\section{Introduction}\label{sec:INTRO}

The past few years have seen significant advances in our understanding of the physics of black holes. Recent computations of the Page curve \cite{Penington:2019npb, Almheiri:2019psf, Almheiri:2019hni} have shed light on the black hole information paradox \cite{Polchinski:2016hrw} and led to a concrete realization of black hole complementarity \cite{Susskind:1993if, Kiem:1995iy, Banks:2001yp}. Meanwhile, strong evidence has been provided for the absence of global symmetries in quantum gravity \cite{Banks:2010zn, Harlow:2018tng, Harlow:2020bee, Chen:2020ojn, Yonekura:2020ino, Heidenreich:2020pkc, Hsin:2020mfa} and the Weak Gravity Conjecture (WGC) \cite{ArkaniHamed:2006dz, Heidenreich:2016aqi, Montero:2016tif, Lee:2019tst, Lee:2018urn}---two conjectures motivated by demanding consistent black hole decay---as well as the Swampland Distance Conjecture (SDC) \cite{Ooguri:2006in, Grimm:2018ohb, Heidenreich:2018kpg, Blumenhagen:2018nts, Corvilain:2018lgw, Grimm:2018cpv, Gendler:2020dfp}. The consequences of these conjectures have been studied extensively.  

At the same time, there has been a revival of interest in questions about de Sitter space and inflation, though as of yet these questions have not been clearly answered. Various works have considered bounds on scalar field potentials in quantum gravity \cite{Obied:2018sgi, Garg:2018reu, Ooguri:2018wrx, Andriot:2018mav, Rudelius:2019cfh, Bedroya:2019snp}, called into question the existence of de Sitter vacua in quantum gravity \cite{Dvali:2017eba, Obied:2018sgi, Danielsson:2018ztv}, and advocated for alternative models of dark energy \cite{Agrawal:2018own, Brandenberger:2018xnf, Agrawal:2019dlm, Anchordoqui:2019amx}. Yet at the same time, other works have made progress in putting claimed de Sitter constructions in string theory on more solid footing \cite{Moritz:2017xto, Akrami:2018ylq, Hamada:2019ack, Kachru:2019dvo, Dasgupta:2019vjn, Hamada:2021ryq}. Similarly, various works have proposed constraints on inflation in quantum gravity \cite{Rudelius:2015xta, Montero:2015ofa, Brown:2015iha, Heidenreich:2015wga, Buratti:2018xjt, Agrawal:2018own}, while other works have attempted to evade these constraints \cite{Bachlechner:2015qja, Hebecker:2015rya, Achucarro:2018vey}.

The goal of this paper is to draw parallels between the physics of black holes and the physics of de Sitter cosmologies, which ideally may allow us to translate some of the recent lessons learned in the former area into progress in the latter. At the same time, we will also highlight connections between semiclassical analyses of black holes and cosmology and recent developments in the ``Swampland program.'' In particular, we will argue that older studies on the difficulties of defining asymptotic observables in an expanding universe may shed light on recent studies of scalar field potentials in string theory, and in turn, recent progress on the Weak Gravity Conjecture and black hole complementarity may shed light on asymptotic observables in de Sitter space.

The remainder of this paper is structured as follows. In Section \ref{sec:Asymptotic}, we review the difficulties of defining asymptotic observables in expanding spacetimes. In Section \ref{sec:Census}, we review the difficulties of defining asymptotic observables, and making predictions, in an eternally inflating cosmology. In Section \ref{sec:Scalar}, we connect the problem of defining asymptotic observables to the de Sitter Conjecture \cite{Obied:2018sgi}. We advocate for a particular ``strong form'' of the de Sitter Conjecture, recently proposed by the author in \cite{Rudelius:2021oaz}, and we explain why this Strong de Sitter Conjecture may be thought of as a sort of Weak Gravity Conjecture \cite{ArkaniHamed:2006dz}. In Section \ref{sec:SDC}, we connect the thermal fluctuations that occur in quintessence models to the Swampland Distance Conjecture \cite{Ooguri:2006in}. In Section \ref{sec:Decay}, we emphasize that the experiences of observers in an eternally inflating cosmology are dictated by whether or not domain walls satisfy a version of the Weak Gravity Conjecture, which suggests a parallel between black hole decay (which involves the ordinary Weak Gravity Conjecture for charged particles) and de Sitter vacuum decay (which involves this version of the Weak Gravity Conjecture for domain walls). We conclude with a discussion of our results and remaining questions in Section \ref{sec:DISC}.


\section{Asymptotic Observables in an Expanding Universe}\label{sec:Asymptotic}

To date, all precise formulations of quantum gravity involve either Anti-de Sitter (AdS) spacetimes or asymptotically flat spacetimes. This is related to the fact that such spacetimes allow us to define asymptotic observables: in the former, such observables are represented by correlation functions of a conformal field theory living at the boundary of AdS, though the AdS/CFT correspondence. In the latter, observables are represented by S-matrix elements.

In cosmology, on the other hand, it is not nearly so simple to define asymptotic observables. In many cosmologies, the presence of an initial singularity precludes the existence of an S-matrix, though this issue may perhaps be sidestepped by assuming a unique initial state and writing down an S-vector that describes only the final state amplitudes \cite{Witten:2001kn, Hellerman:2001yi}. However, an even larger issue looms: as emphasized in e.g. \cite{Bousso:2004tv, Susskind:2007pv}, FRW cosmologies do not have a property known as ``asymptotic coldness'': the energy density of these spacetimes does not tend to zero at spatial infinity on a fixed Cauchy slice, and relatedly fluctuations of the geometry extend indefinitely. Any observer looking into the past at finite time can only access a finite portion of such a universe, whereas the unobserved region of space contains an infinite amount of energy and (perhaps) an infinite amount of information. This prevents the observer from accessing the global state of the universe no matter how long they wait, and it precludes an S-matrix or S-vector description.

It is possible that exact asymptotic observables simply do not exist in cosmology, and trying to define them is nothing but a fool's errand. But on the other hand, it is possible that the above issues could be partially circumvented, and some sort of asymptotic observables could be defined even in the absence of asymptotic coldness. The prospects for this depend strongly on the type of cosmology under consideration. Following \cite{Bousso:2004tv}, we review three such possibilities: de Sitter space (with equation of state parameter $w = -1$), Q-space ($-1 < w < -1/3$), and a decelerating universe ($w> -1/3$).

For future reference, let us write down the metric of a (spatially flat) 4d FRW cosmology:
\begin{equation}
ds^2 = - dt^2 + a(t)^2 \left(  dr^2 + r^2 d \Omega_2^2 \right)\,,
\end{equation}
where $a(t)$ is the scale factor, which we assume evolves according to
\begin{equation}
a(t) = \left\{ \begin{array}{cc}
t^{\frac{2}{3(w+1)}} & w > -1 \\
e^{H t} & w = -1
\end{array} \right..
\end{equation}
Here we define the Hubble parameter $H := \dot a/a$, which is constant for $w = -1$ but vanishes in the $t \rightarrow \infty$ limit for $w > -1$. The energy density is then given by $\rho = H^2/3$, and the pressure is given by $p = w \rho$. Throughout this paper we set $8 \pi G = 1$.

\subsection{De Sitter}

\begin{figure}
\centering
\includegraphics[width=40mm]{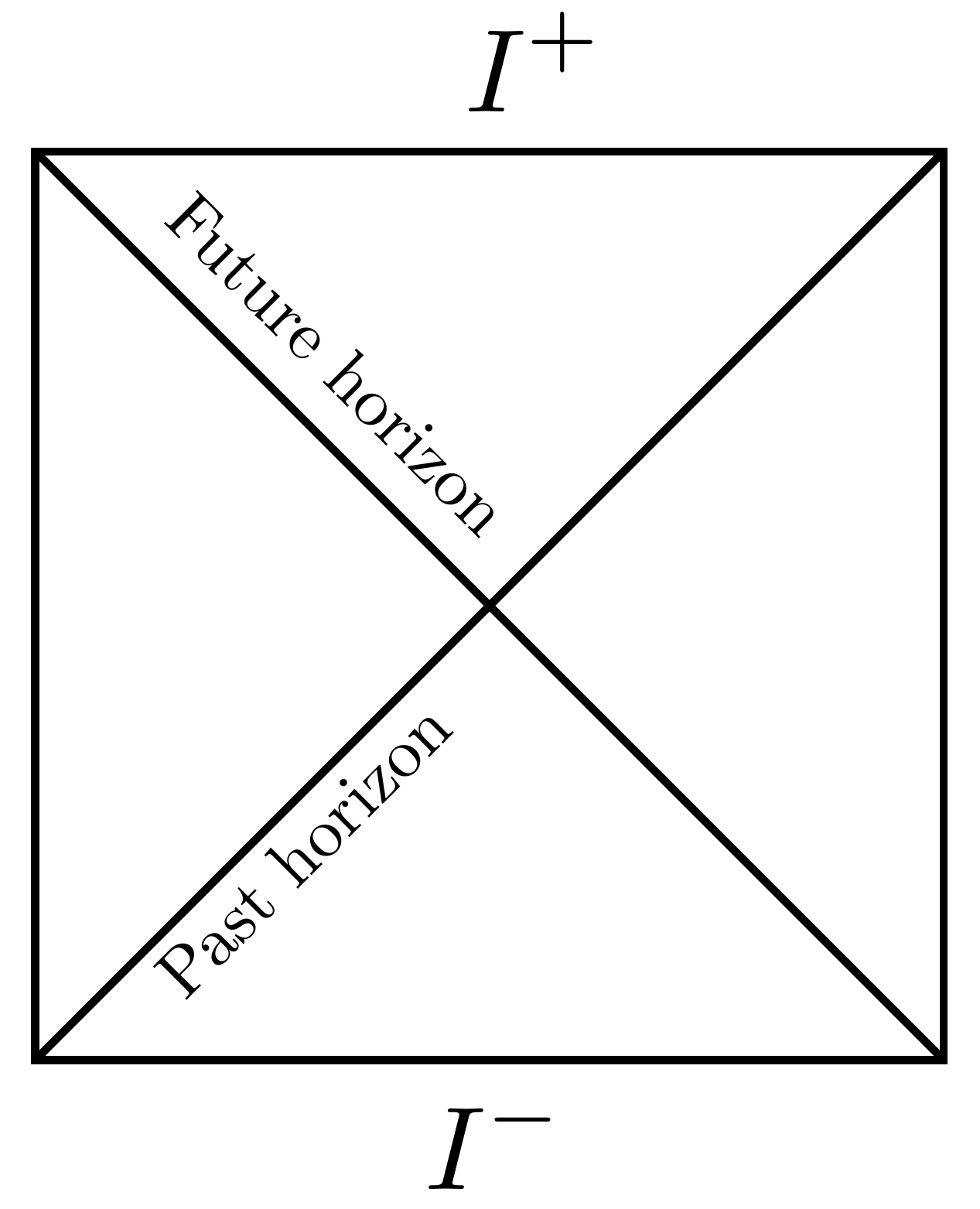}
\caption{The Penrose diagram of de Sitter space.}\label{dSPenrose}
\end{figure}

De Sitter space (dS) has equation of state parameter $w = -1$: the scale factor $a(t)$ expands exponentially at a constant rate for all time. The maximal extension of de Sitter has the metric
\begin{equation}
ds^2 = \frac{1}{H^2 \sin^2 \eta} \left( - d \eta^2 + d \chi^2 + \sin^2 \chi d \Omega_2^2\right)\,.
\end{equation}
and its Penrose diagram is shown in Figure \ref{dSPenrose}. The spatial sections at constant $\eta$ are 3-spheres, so there is no notion of spatial infinity or null infinity: the only asymptotic regions are $I^\pm$. A comoving observer has both past and future horizons of radius $H^{-1}$. The region limited by these respective horizons is called the causal diamond.

Several factors preclude the existence of asymptotic observables in de Sitter space. To begin, de Sitter space has a future horizon: there are regions of spacetime that are forever out of contact of a comoving observer, so no observer can ever witness the final state of the entire universe. Second, de Sitter has a finite entropy,
\begin{equation}
S_{dS} =  \frac{8 \pi^2}{ H^{2}}  \,,
\end{equation}
 and the Hilbert space of quantum gravity in de Sitter is (likely) finite-dimensional, which puts a limit on the complexity of any sort of measurement apparatus that could exist in de Sitter space \cite{Witten:2001kn}. Finally, as noted in \cite{Bousso:2004tv}, de Sitter has a constant temperature 
 \begin{equation}
 T_{\text{dS}} = \frac{H}{2 \pi} \,,
 \end{equation}
  and the resulting thermal fluctuations, if sufficiently energetic, may destroy any observers. Such high-energy thermal fluctuations are Boltzmann-suppressed by a factor of $\exp(- E/ T_{\text{dS}})$, but if one waits long enough such fluctuations are bound to occur, and thus no observers can exist eternally in de Sitter space.
  
 For practical purposes in our own universe, these issues are not so important: the de Sitter entropy is enormous, of order $10^{123}$, and the temperature is minuscule, of order $10^{-61}$, so observables can be defined to very good approximation. In addition, one might also entertain the possibility of ``meta-observables'' \cite{Witten:2001kn}, which are exact quantities that could be determined by a ``meta-observer'' who has access to the entirety of $I^+$ and $I^-$ (see Figure \ref{dSPenrose}). However, neither of these approaches quite addresses the conceptual problem at hand:
 no observer living in de Sitter space can perform a measurement with arbitrary precision, so exact observables do not exist in de Sitter.

\subsection{Q-space}

\begin{figure}
\centering
\includegraphics[width=40mm]{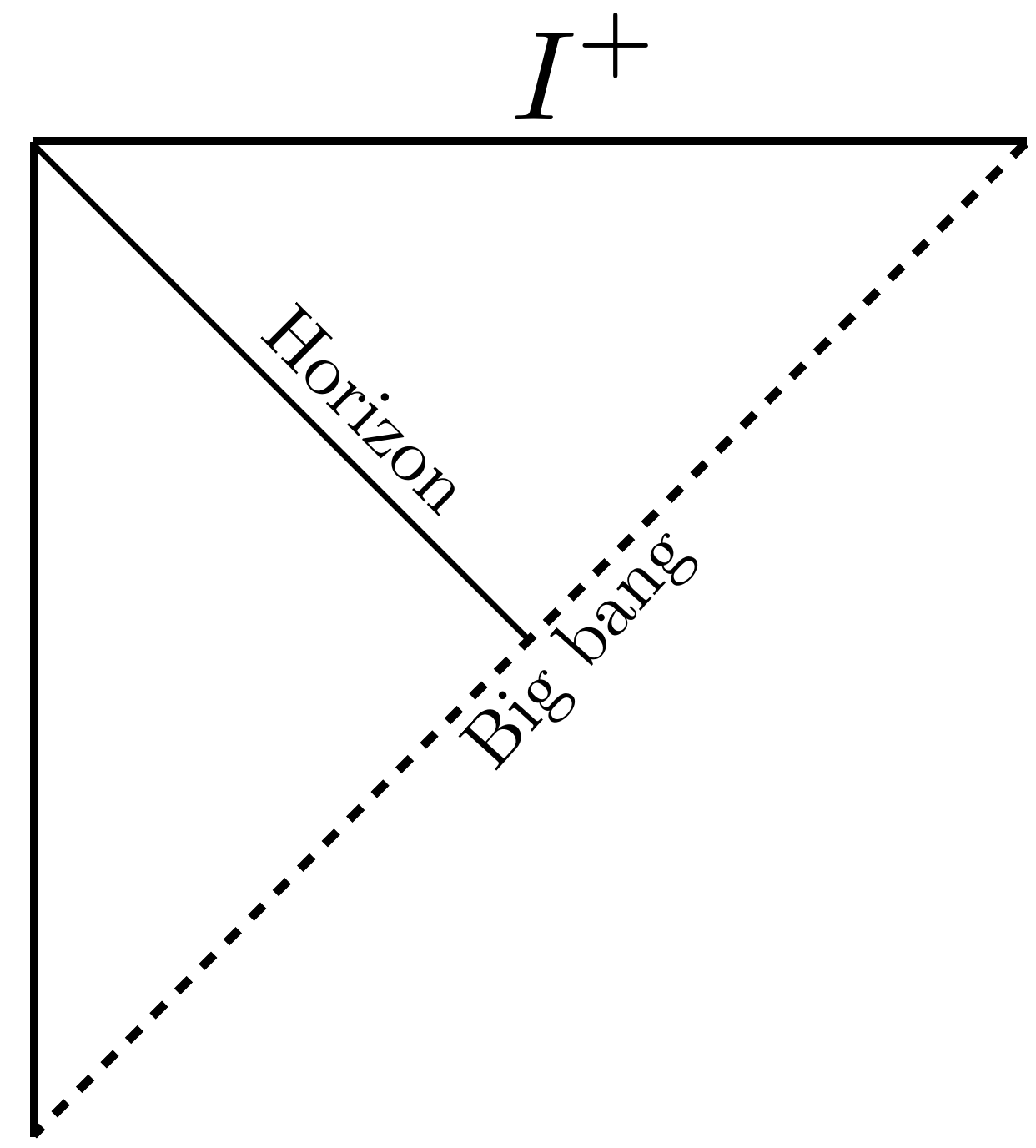}
\caption{The Penrose diagram of Q-space.}\label{QPen}
\end{figure}

Q-space ($-1 <w < -1/3$) describes a quintessence-dominated universe, which accelerates indefinitely into the future but with decreasing Hubble parameter $H=\dot a/a$. Its associated Penrose diagram is shown in Figure \ref{QPen}. An observer's causal diamond is bounded by a light-like big bang singularity and a future event horizon of radius 
\begin{equation}
R_E = -\frac{3(w+1)}{3w +1} t = - \frac{2}{3w + 1}   H^{-1}\,.
\end{equation} 
There is also an apparent horizon of radius
\begin{equation}
R_A =  H^{-1} = \frac{3}{2} (w+1) t \,.
\end{equation} 
Note that the size of these horizons grows without bound in the limit $t \rightarrow \infty$, and correspondingly there is no bound on the amount of entropy that can exist within a comoving observer's causal diamond. Indeed, such an observer will experience a thermal heat bath with temperature 
\begin{equation}
T_{Q} = \frac{H}{2 \pi} = \frac{3}{ \pi (w+1) t} \,,
\end{equation}
 and although this temperature decreases indefinitely with time, arbitrarily high-entropy fluctuations will occur even at late times. However, unlike in de Sitter, the Boltzmann suppression of high-energy fluctuations increases with time, and shortly after the Hubble scale $H$ drops below a given energy $E$, the probability of ever again observing a thermal fluctuation of energy $E$ drops quickly to zero \cite{Bousso:2004tv}.

Thus, life is not quite as bad for an observer in Q-space as it is for an observer in de Sitter: although there is still a future cosmic horizon, there is no limit on the entropy contained within such a horizon, and correspondingly there is no inherent limit on the precision of a measurement. Furthermore, observers in Q-space are rarely destroyed by high-energy thermal fluctuations, whereas in de Sitter this will eventually occur with probability 1.

Nonetheless, an important problem faces any observer in Q-space who wants to measure some observable with ever-greater precision: although the maximal entropy allowed within the horizon may increase indefinitely, the only source of such entropy is the thermal fluctuations of massless (or nearly massless---see Section \ref{sec:SDC}) particles from the horizon. Even in a more realistic cosmology with a period of matter/radiation domination preceding accelerated expansion, the entropy of conventional matter/radiation that enters the causal diamond of the observer by classical evolution is finite: eventually, a pair of comoving objects will exit one another's horizons and cease to interact. Thus, constructing an experimental apparatus of arbitrarily large complexity would require the observer to somehow harness the entropy of thermal radiation from the horizon, and it is not clear that this is possible \cite{Bousso:2004tv}.

In summary, while the prospects for defining asymptotic observables are more promising in Q-space than in de Sitter, there are nonetheless some important difficulties that must be overcome. We will now see that some of these difficulties are avoided in a decelerating universe.

\subsection{Decelerating Universes}

\begin{figure}
\centering
\includegraphics[width=40mm]{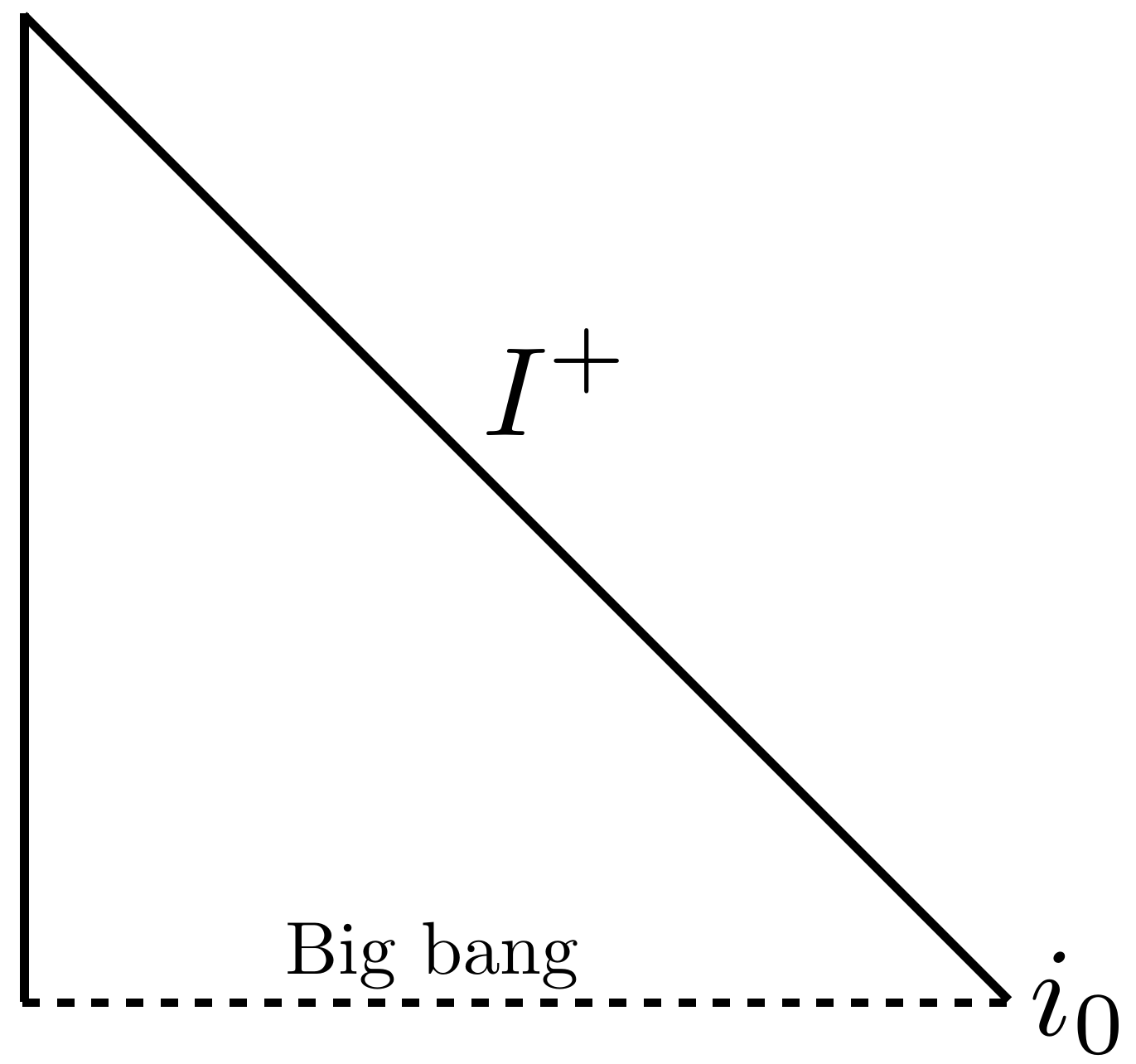}
\caption{The Penrose diagram of a decelerating FRW cosmology.}\label{decelP}
\end{figure}

The Penrose diagram of a decelerating universe $w > -1/3$ is shown in Figure \ref{decelP}: essentially, it is obtained by turning the Penrose diagram of Q-space upside down. A decelerating universe does not have a future horizon: eventually, any two coming objects will enter one another's past light cones, and correspondingly the entropy accessible to a given observer diverges in the asymptotic future. There is therefore no limitation on the complexity of a possible experimental apparatus, and there is no limit on the size of space visible to a late-time observer. The theory at late times approaches that of particles interacting in flat space: in short, the conditions for observers are as ideal as they could be in a universe without asymptotic coldness.


\section{Asymptotic Observables in Eternal Inflation}\label{sec:Census}

Our universe is experiencing accelerated expansion, and thus far dark energy has no observed departure from $w=-1$. This is difficult to square with the aforementioned difficulties facing de Sitter space (among other difficulties \cite{Dyson:2002pf, Goheer:2002vf, Page:2006nt}), but string theory may point us in the right direction: in string theory, de Sitter vacua are generally expected to be metastable, decaying via bubble nucleation to other vacua in the string Landscape \cite{Coleman:1980aw}. These vacua may in turn decay via bubble nucleation, and this process continues until a terminal vacuum is reached \cite{Susskind:2012pp,Harlow:2011az}: presumably either a $\Lambda < 0$ AdS vacuum (which promptly crunches) or a $\Lambda = 0$ vacuum. In the latter case, the vacuum in question may lie in an asymptotic limit of scalar field space, in which case bubble nucleation may produce either an accelerating or decelerating universe, depending on the shape of the potential---we will elaborate on the details of this point in the next section.

The upshot of this is that the difficult task of defining asymptotic observables in de Sitter space has been exchanged for the difficult task of defining asymptotic observables in an eternally inflating cosmology.\footnote{A number of works have questioned the plausibility of eternal inflation \cite{Banks:2003pt, ArkaniHamed:2007ky, ArkaniHamed:2008ym, Olum:2012bn, Dvali:2017eba, Rudelius:2019cfh, Bedroya:2020rac, Seo:2021bpb}, though at present there is no sharp no-go result which forbids it.} In such a cosmology, anything that \emph{can} happen \emph{does} happen an infinite number of times, raising the difficult question of how to define a measure on the Landscape of all possible universes. For a nice review of the measure problem in eternal inflation, see \cite{Freivogel:2011eg}.

One approach to the measure problem, notably advocated by Susskind \cite{Susskind:2007pv}, is to introduce a comoving observer who looks back into the past and collects data. The expected observations of this observer then define a measure on the Landscape: namely, the probability $p_i$ of outcome $i$ is given in terms of the number of measurements $N(t)$ in the past light cone of the observer at time $t$ and the number of times $N_i(t)$ the observer measures $i$ by \cite{Bousso:2011up}
\begin{equation}
p_i := \lim_{t \rightarrow \infty} \frac{N_i(t)}{N(t)}\,.
\end{equation}
Such an observer is called a ``census taker.''

Of course, there are different possible census takers one could choose, i.e., different locations where the ``census bureau'' could be located. Some census takers will end their lives in an AdS vacuum, which ends with a singular crunch. Such a census taker only has access to a finite amount of entropy before meeting their demise, so they cannot measure any observable to arbitrary precision. Other census takers may end up in a bubble of Q-space, in which the acceleration of the universe asymptotes to zero. This is a much more promising location for a census bureau, but the challenges of defining asymptotic observables here are just as thorny here as they were in Q-space: in particular, a pair of comoving objects will eventually exit one another's horizons, and no observer can see all the way to spatial infinity. 

\begin{figure}
\centering
\includegraphics[width=80mm]{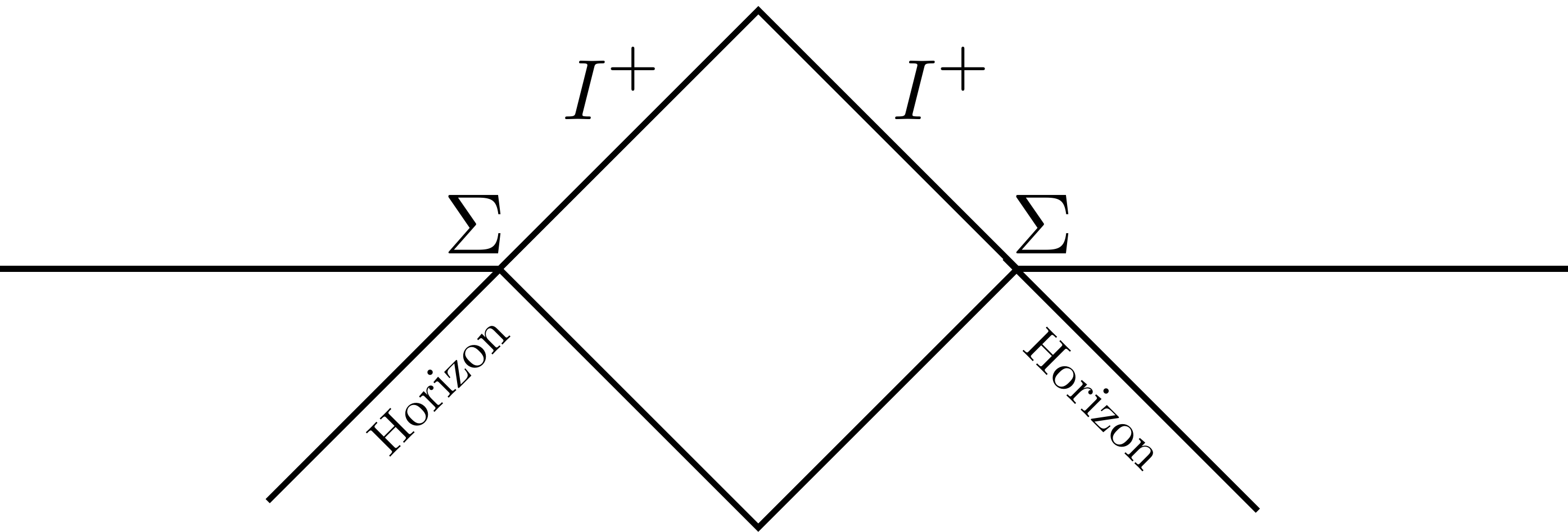}
\caption{A hat region of an eternally inflating cosmology.}\label{EI}
\end{figure}

The most promising location for a census bureau is in a ``hat region'': a bubble containing a decelerating universe, so named because of the hat-like feature it introduces to the Penrose diagram (see Figure \ref{EI}). Such a region has many of the nice features of a decelerating universe: a census taker in the hat region region will observe an infinite amount of entropy if she waits long enough, any two comoving objects in the hat region will eventually come into causal contact, and physics at late times can be described by an effective field theory in flat space.

Some preliminary but noteworthy progress has been made in understanding the physics in such regions. In particular, \cite{Freivogel:2006xu} proposed that the physics in a hat region has a dual description in terms of a CFT coupled to Liouville gravity living at spatial infinity $\Sigma$ (see Figure \ref{EI}). In \cite{Harlow:2010my}, Harlow and Susskind further conjectured that the maximum precision of a dual description of a cosmological geometry is determined by the maximal entropy bound on the past light cone of a census taker, and they argued that this description can be exact, i.e., ultraviolet complete, only if there exists a census taker with an infinite entropy bound (also known as a ``maximal'' census taker). In principle, a census taker in a bubble of Q-space fits this description as well, since they too have an infinite entropy bound, but to date there has been little to no work in understanding the dual descriptions of such universes.

One issue facing a census taker in hat region, which is absent in the case of a decelerating FRW universe, is the existence of a cosmic horizon: as shown in Figure \ref{EI}, such a census taker can see all the way to spatial infinity $\Sigma$ in her own bubble, but there are other parts of spacetime hidden beyond the cosmic horizon. This issue, however, is probably not devastating from the perspective of defining asymptotic observables: indeed, black hole horizons exist in asymptotically AdS spacetimes. Here, a crucial role is played by the notion of black hole complementarity \cite{Susskind:1993if, Kiem:1995iy, Banks:2001yp}: the physics of the black hole interior is encoded in the radiation emitted by the black hole horizon. It is plausible that a similar notion of cosmic horizon complementarity is relevant here, so that the global structure of the eternally inflating spacetime is encoded in the radiation emitted from the cosmic horizon \cite{Susskind:2007pv, Sekino:2009kv}, ostensibly allowing the census taker to define a global measure on the spacetime beyond her horizon \cite{Bousso:2011up}. Similarities between de Sitter cosmic horizons and black hole horizons have recently been investigated in \cite{Aalsma:2020aib,Aalsma:2021bit, Aalsma:2021kle}.


\section{Asymptotic Observables and the de Sitter Conjecture}\label{sec:Scalar}

While the difficulties of constructing de Sitter vacua in string theory are not new, they have drawn renewed interest since the publication of what is now referred to as the ``de Sitter Conjecture (dSC)'' \cite{Obied:2018sgi}. In this work, the authors proposed a bound on the gradient of scalar field potentials $V$ in quantum gravity in $d$ spacetime dimensions of the form
\begin{equation}
|\nabla V| \geq c_d V \,,
\label{eq:dSC}
\end{equation}
where $c_d$ is an $O(1)$ constant in $d$-dimensional Planck units to be determined later. This bound is almost surely violated at the maximum of the standard model Higgs potential, so it is very unlikely that it applies universally in quantum gravity \cite{Denef:2018etk}. However, it is quite plausible that a bound of this form applies in asymptotic regions of scalar field space \cite{Dine:1985he, Ooguri:2018wrx}, and indeed many examples in string theory satisfy this criterion \cite{Wrase:2010ew, Obied:2018sgi, Andriot:2019wrs}. In the following section, we will review one general argument for the validity of this bound in asymptotic regions of scalar field space, originally given in \cite{Hebecker:2018vxz}.

In order to determine the domain of validity of the dSC bound \eqref{eq:dSC}, it is crucial to fix the precise value of the $O(1)$ constant $c_d$, and relatedly to understand the physical principle underlying this bound. As an analogy, the precise $O(1)$ coefficient $\gamma_d$ appearing in the WGC bound $|q|/m \geq \gamma_d$ is fixed by the physical principle that non-supersymmetric black holes must be able to decay. We would like a similar statement here to fix the value of $c_d$.

\subsubsection{Review: Sharpening the dSC}

In \cite{Rudelius:2021oaz}, it was noted that the dSC is exactly preserved under dimensional reduction if we set 
\begin{equation}
c_d = \cdstrong:= \frac{2}{\sqrt{d-2}}\,.
\end{equation}
In other words, a theory which saturates the bound \eqref{eq:dSC} with this value of $c_{d}$ will still saturate the bound after dimensional reduction. Of course, it is not obvious that \emph{every} valid Swampland conjecture must be exactly preserved under dimensional reduction, but it is worth noting that some of the most well-supported Swampland conjectures are, including the Weak Gravity Conjecture and the absence of global symmetries \cite{Heidenreich:2015nta, Heidenreich:2019zkl}.

The value $c_d = \cdstrong$ is also special from the perspective of the late-time expansion of the universe: a scalar field rolling in a potential of the form 
 \begin{equation}
 V \propto \exp ( - \lambda \phi )
 \end{equation}
  will drive expansion of the universe with equation of state 
  \begin{equation}
  w =-1 + \frac{1}{2} \frac{d-2}{d-1} \lambda^2\,,
  \end{equation}
   which for $\lambda > \cdstrong$ implies 
   \begin{equation}
   w \geq - \frac{d-3}{d-1} \,.
   \end{equation}
    This is precisely the strong energy condition in $d$ dimensions, which is precisely the condition that forbids accelerated expansion of the universe. Consequently, we define the ``Strong de Sitter Conjecture'' as the statement that the strong energy condition should be satisfied at late times\footnote{The ``late times'' condition is necessary to exclude the possibility that a scalar field in an asymptotic region is given a kick of kinetic energy so that it violates the strong energy condition for a brief time. We thank Thomas Van Riet for pointing out this possibility to us.} in asymptotic limits of scalar field space, which is equivalent to the de Sitter bound \eqref{eq:dSC} with $c_d = \cdstrong := 2/\sqrt{d-2}$ if we suppose that the energy density is dominated by the scalar field condensate. Note that we do not expect this condition to hold outside of such asymptotic regions of scalar field space: indeed, the strong energy condition is violated even in our own universe.
    
\subsubsection{Supersymmetry and the Strong dSC}

String vacua with vanishing vacuum energy $V = 0$ are expected to be supersymmetric, except in certain asymptotic limits of scalar field space \cite{Harlow:2010my} such as the weak coupling limit of non-supersymmetric, $SO(16) \times SO(16)$ heterotic string theory \cite{AlvarezGaume:1986jb}.\footnote{We thank Irene Valenzuela for explaining this example to us.} Many string vacua in asymptotic limits of scalar field space \emph{are} supersymmetric, however, and in this context compelling evidence for the Strong dSC was given in \cite{Hellerman:2001yi}: the authors of that paper argued that within an accelerating cosmology in four dimensions, a scalar field cannot asymptote to a zero-energy supersymmetric minimum, and their argument can be extended trivially to general spacetime dimensions. In particular, the potential around a stable supersymmetric minimum in $d$ dimensions must take the form \cite{Townsend:1984iu, Skenderis:1999mm}:
\begin{equation}
V(\phi^i) = 2 (d-2)  \left( (d-2) (\nabla W)^2  - (d-1)  W^2 \right) \,,
\end{equation}
where $W$ is the superpotential. If we then assume that the potential takes the asymptotic form $V \propto \exp( - \lambda \phi)$ as $\phi \rightarrow \infty$, we must impose $W \approx W_0 \exp(- \lambda \phi/2)$, which leads to 
\begin{equation}
V(\phi) \approx 2 (d-2) W_0^2 e^{-  \lambda \phi} \left( (d-2) \frac{\lambda^2}{4}  - (d-1) \right) \,.
\end{equation}
This is positive only if 
\begin{equation}
\lambda > 2 \sqrt{\frac{d-1}{d-2}} \,,
\end{equation}
 which immediately implies $\lambda > 2/ \sqrt{d-2}$, thereby satisfying the Strong dSC.

 \subsubsection{Asymptotic Observables and the Strong dSC}

From the discussion in the previous section, we see that the Strong dSC may also be motivated by a more physical principle: the existence of asymptotic hat regions, which may be necessary for the existence of well-defined observables in spacetimes with positive vacuum energy, requires not only that $V \rightarrow 0$ limits exist in scalar field space, but also that these limits occur in regions with decelerating expansion. Conversely, the difficulty of defining exact observables in asymptotic Q-space may signal an inconsistency with such spacetimes, implying that a scalar field potential which violates the Strong dSC bound in asymptotic regions of scalar field space must reside in the Swampland. Note that the existence of $V \rightarrow 0$ regions with decelerating expansion is uncontroversial: the existence of $\mathcal{N}=2$ Minkowski vacua in string theory is beyond dispute. The radical suggestion here is the converse: the nonexistence of asymptotic Q-space in string theory would be surprising.

In a decelerating universe (in contrast to Q-space), physics at late times is governed by an effective theory of interacting particles in flat space.
In a sense, the Strong dSC should therefore be thought of as a type of Weak Gravity Conjecture: just as the ordinary WGC holds that gravity can be decoupled from electromagnetism at low energies and the scalar WGC \cite{Palti:2017elp} holds that gravity can be decoupled from interactions mediated by a scalar field at low energies, the Strong dSC holds that gravitational expansion of the universe can be decoupled from other interactions at late times.

 \subsubsection{String Theory and the Strong dSC}
 
Many examples in string theory satisfy the Strong dSC, such as the KKLT scenario and the LVS scenario. More examples are discussed in \cite{Rudelius:2021oaz}.

Here, we instead focus on a pair of caveats: apparent counterexamples to the Strong dSC in string theory. The first class of apparent counterexamples come from string compactifications which saturate the ``Transplanckian Censorship Conjecture'' bound \cite{Bedroya:2019snp}:
\begin{equation}
|V'| \geq \frac{2}{\sqrt{(d-1)(d-2)}} V \,.
\label{TCCbound}
\end{equation}
String compactifications saturating this bound were studied in e.g.\cite{Wrase:2010ew,Andriot:2019wrs,Andriot:2020lea}. Since the coefficient on the right-hand side of this inequality is smaller than $\cdstrong$, these examples naively seem to violate the Strong dSC.

However, as previously discussed in \cite{Rudelius:2021oaz}, this conclusion is premature: the expression $V'$ appearing in \eqref{TCCbound} is the derivative of the potential with respect to the geodesic distance along certain geodesics in field space. The expression $\nabla V$ appearing in the Strong dSC is instead the gradient of the potential, which receives contributions from all the scalar fields in the theory. Once these additional contributions are included, it is conceivable that the Strong dSC may be satisfied even while \eqref{TCCbound} is saturated. Indeed, one can check that this happens in e.g. heterotic string theory compactified to four dimensions \cite{Rudelius:2021oaz}.

Finally, let us note a second class of important examples, which arise from compactifications of supercritical string theory.\footnote{We are grateful to Miguel Montero and Irene Valenzuela, as well as Eva Silverstein, for independently bringing this example to our attention.} Supercritical string theories actually saturate the Strong dSC bound \cite{Hellerman:2006nx}, but certain compactifications of supercritical string theories violate the bound and give rise to accelerated expansion \cite{Dodelson:2013iba, Ellis:2020nnp}. However, this accelerated expansion lasts for only a finite time before perturbative control is lost, so these cosmologies do not contradict the Strong dSC, which by definition forbids accelerated expansion only at arbitrarily late times in asymptotic limits of scalar field space.


\section{Thermal Fluctuations and the Swampland Distance Conjecture}\label{sec:SDC}

We have argued that the Strong dSC value $\cdstrong = 2/\sqrt{d-2}$ is a particularly natural choice for the $O(1)$ constant $c_d$ appearing in the dSC bound, motivated both by dimensional reduction and by the improved prospects for defining asymptotic observables in a decelerating universe. Nonetheless, it is worthwhile to consider the alternative possibility of an observer who ends their life in Q-space.

Recall from our discussion above that the prospects for asymptotic observables in Q-space are not completely hopeless: perhaps the greatest difference between a census taker in Q-space and a census taker in a hat region is that the Q-space observer cannot access arbitrarily large amounts of information through classical evolution: the only way to build a measuring device of arbitrary complexity is through low-energy, high-entropy thermal fluctuations coming from the cosmic horizon. To assess the possibility of asymptotic observables in Q-space, we must therefore investigate the thermal fluctuations that occur in such cosmologies.

At late times, the temperature of the future horizon in Q-space tends to zero, so the thermal fluctuations become softer and softer. Eventually, one expects that only massless modes will be emitted \emph{unless} there are states whose energy decreases appropriately with time. Indeed, the existence of such states is a stipulation of another well-tested Swampland conjecture called the Swampland Distance Conjecture (SDC) \cite{Ooguri:2006in},\footnote{Strictly speaking, the Swampland Distance Conjecture applies only to massless scalar field moduli in supersymmetric theories, whereas we are concerned here with a ``refined version'' of the conjecture, which holds that this conjecture should apply in asymptotic limits of scalar field space in non-supersymmetric theories as well \cite{Klaewer:2016kiy}. Compelling, general arguments for both the SDC and its refinement have been given in \cite{Grimm:2018ohb, Heidenreich:2018kpg}, and we will not distinguish the two conjectures in this work.} which holds that in any asymptotic limit $\phi \rightarrow \infty$ in scalar field space, there must exist a tower of particles, each labeled by a positive integer $n$, whose masses scale with $\phi$ and $n$ as
\begin{equation}
m_n \sim n m_0 e^{- \alpha \phi}\,,
\end{equation}
for $m_0$ an arbitrary mass scale and $\alpha$ some $O(1)$ constant in Planck units. Like the value of $c_d$ appearing in the dSC bound \eqref{eq:dSC}, the question of what values of $\alpha$ are allowed in quantum gravity is a topic of ongoing research.

Motivated by the SDC, let us now consider thermal emission from the Q-space horizon in four dimensions. As discussed in \cite{Bousso:2004tv}, the probability of thermal emission per unit time of a state of energy $E$ is given roughly by
\begin{equation}
P_E \sim \frac{1}{R_A} \exp\left[S(E) - 2 \pi E R_A \right]\,,
\label{PE}
\end{equation}
where $R_A  =H^{-1}$, $\exp(S(E))$ is the number of states of energy $E$, and we are assuming that there are only $O(1)$ states of energy below the Q-space temperature $T_{Q}  = H/(2 \pi)$. According to Bousso's D-bound \cite{Bousso:2000md}, which generalizes the Bekenstein bound \cite{Bekenstein:1972tm, bekenstein:1980jp} to de Sitter space, the term in the exponent in \eqref{PE} is necessarily negative.

Integrating this result over all time $t > t_0$, we find the probability of a thermal fluctuation of energy $E$ after time $t_0$:
\begin{align}
\mathcal{P}(E) &=  \int_{t_0}^\infty dt P_E(t) \nonumber \\
&= \int_{t_0}^\infty dt \frac{1}{R_A(t)} \exp\left[S(E) - 2 \pi E R_A \right]\,.
\end{align}
For constant $E \gg R_A^{-1}(t_0)$, the Boltzmann suppression $\exp(- 2 \pi E R_A(t_0))$ dominates, and the total probability $\mathcal{P}(E)$ is small. However, in a theory satisfying the SDC, the energy of a single-particle state in the tower itself changes with time, $E = E(t)$. In particular, if we have
\begin{equation}
m_n \sim n m_0 e^{- \alpha \phi}\,,~~~V(\phi) \sim V_0 e^{-\lambda \phi}\,,
\end{equation} 
for $\phi$ a canonically normalized scalar field, then the energy of such a state behaves as $E \sim E(t_0) \exp (- \alpha \phi) $, whereas the radius $R_A$ scales as $R_A \sim R_A(t_0) \exp(\lambda \phi /2)$. This means that if 
\begin{equation}
\alpha = \frac{\lambda}{2}\,,
\end{equation} 
then the Boltzmann suppression factor $\exp(- 2 \pi E R_A)$ remains roughly constant over time, and the total probability is bounded from below by
\begin{align}
\mathcal{P}(E) &\geq  \int_{t_0}^\infty dt \frac{1}{R_A(t)} \exp\left[- 2 \pi E(t) R_A(t) \right]  \nonumber \\
& \gtrsim \frac{2}{3 (w+1)} \exp\left[- 2 \pi E(t_0) R_A(t_0) \right]  \int_{R_A(t_0)}^\infty \frac{d R_A}{R_A}  \,,
\end{align}
where we have used the fact that $R_A = \frac{3 (w+1)}{2} t$. This integral diverges logarithmically, so we see that thermal emission of particles in the SDC tower continues indefinitely when $\alpha \geq \lambda/2$. 

Let us make a few remarks on how this result fits with existing literature. Reference \cite{Andriot:2020lea} proposed precisely the relationship $\alpha \geq c_d/2$ between the exponent of the SDC tower $\alpha$ and the coefficient of the dSC bound \eqref{eq:dSC}, based on the fact that many examples in string theory seem to satisfy or saturate this bound \cite{Andriot:2020lea, Grimm:2018ohb} if one sets
\begin{equation}
c_d = \cdTCC := {2 \over \sqrt{(d-1)(d-2)}} \,,
\label{eq:TCC}
\end{equation}
which is the value of $c_d$ specified by the ``Transplanckian Censorship Conjecture (TCC)'' \cite{Bedroya:2019snp}.
We see here that the bound $\alpha \geq \lambda/2$ follows from the requirement that states in the SDC tower are thermally produced at late times, which together with the dSC bound $\lambda \geq c_d$ implies $\alpha \geq c_d/2$. Whether or not such a requirement should hold---and whether or not asymptotic Q-space can even occur in quantum gravity---is a topic for future research.

Under the assumption of the SDC, \cite{Hebecker:2018vxz} presented a compelling argument for the validity of the dSC in asymptotic regions of scalar field space: in the limit $\phi \rightarrow \infty$, the SDC implies the existence of a large number of light species $N$, which in turn lead to a UV cutoff on effective field theory that is parametrically below the 4d Planck scale \cite{Veneziano:2001ah, ArkaniHamed:2005yv, Dvali:2007wp,Dvali:2007hz}:
\begin{equation}
\Lambda_{\text{UV}} \sim \frac{M_{\text{Pl}} }{ \sqrt{ N(\Lambda_{\text{UV}} )} }\,.
\end{equation}
According to the SDC, the number of light species is given by
\begin{equation}
N(\Lambda_{\text{UV}} ) \gtrsim \frac{\Lambda_{\text{UV}}}{m_0 e^{-\alpha \phi}}\,,
\end{equation}
so
\begin{equation}
\Lambda_{\text{UV}} \sim e^{- \alpha \phi /3}\,.
\end{equation}
On the other hand, the Hubble scale $H$ acts as an IR cutoff on effective field theory, and this scales with $\phi$ as
\begin{equation}
H \sim V^{1/2} \sim e^{- \lambda \phi/2}\,.
\end{equation}
Requiring that the UV cutoff $\Lambda_{\text{UV}} $ is larger than the IR cutoff $H$ as $\phi \rightarrow \infty$ therefore implies
\begin{equation}
\alpha \leq \frac{3}{2} \lambda\,,
\end{equation}
which is consistent with the condition $\alpha \geq \lambda/2$.

It is worth emphasizing that the relationship $\alpha \geq \lambda/2$ we have found here assumes that we are in Q-space, so the universe is accelerating at late times, i.e., $0 < \lambda < 2/\sqrt{d-2}$. Our calculation tells us nothing about the coefficient $\alpha$ of an SDC tower in a decelerating universe ($\lambda > 2 / \sqrt{d-2}$) or a supersymmetric theory in flat space ($V \equiv 0$), so in particular the bound $\alpha \geq c_d/2$ is not applicable if one assumes the Strong dSC.


\section{Vacuum Decay and the Weak Gravity Conjecture}\label{sec:Decay}

So far, we have seen how the experiences of different observers in an expanding universe point towards the Strong de Sitter Conjecture and perhaps the Swampland Distance Conjecture: census takers have helped shed light on the Swampland. We now present an instance of the opposite: the Weak Gravity Conjecture (WGC) may help shed light on the observations of census takers in a hat region.

Consider a maximal census taker, i.e., a census taker who begins her life in a de Sitter phase and ends her life in a hat region. Such a census taker is maximal in the sense that she will eventually have access to infinite entropy. However, in the standard description of eternal inflation, her past light cone will not contain the entirety of the expanding universe (see Figure \ref{EI}): her observations are limited by a cosmic horizon.
As discussed in Section \ref{sec:Census} above, this issue could be circumvented by a suitable notion of cosmic horizon complementarity, in which the region behind the observer's horizon is somehow encoded in the radiation she receives from this horizon.\footnote{Some support of this idea may come from recent work \cite{Hartman:2020khs}, which found islands in a dS$_2$ spacetime with a hat region, indicating that an observer in the hat region has access to information about the inflating region, analogous to the way islands in black hole spacetimes indicate that an asymptotic observer has access to information about the black hole interior \cite{Penington:2019npb, Almheiri:2019psf, Almheiri:2019hni}. It is unclear at present if this story can be extended to de Sitter spacetimes in more than two dimensions, however.}

\begin{figure}
\centering
\includegraphics[width=80mm]{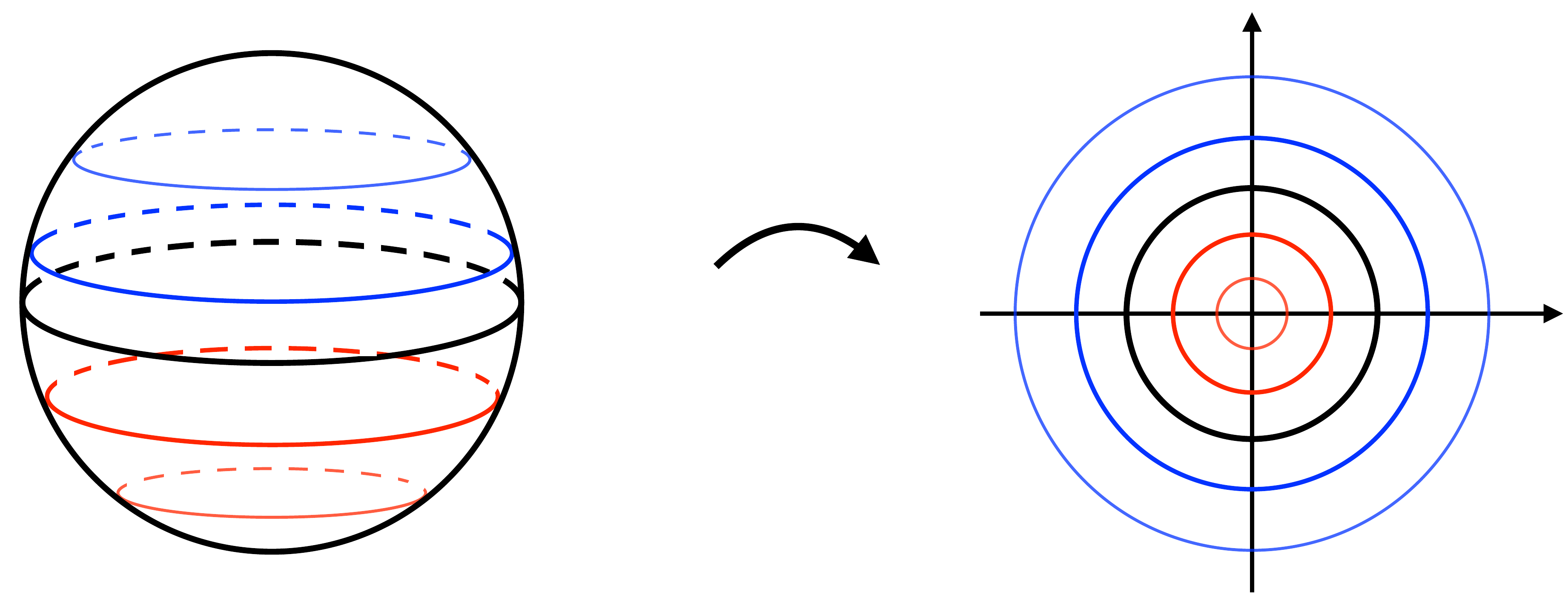}
\caption{Stereographic projection of the 2-sphere. Latitudes in the southern hemisphere map to circles in the interior of the unit disk, and their radius of curvature on the 2-sphere grows with increasing size in the stereographic plane. Latitudes in the northern hemisphere, on the other hand, map to circles outside the unit disk, and the radius of curvature of these latitudes on the 2-sphere actually \emph{decreases} while the size of the associated circles in the plane \emph{increases}.}\label{Stereo}
\end{figure}

Nonetheless, there is a somewhat surprising aspect of this picture: not all decays from de Sitter space into the $\Lambda = 0$ vacuum can occur in the past light cone of a maximal census taker \cite{Freivogel:2011eg}. This is due to the counterintuitive fact that the domain wall separating a $\Lambda_i > 0$ vacuum from a $\Lambda_f = 0$ vacuum may appear to recede from the perspective of \emph{both} the inside observer and the outside observer. (This fact can be visualized by analogy with stereographic projection of the 2-sphere, as shown in Figure \ref{Stereo}.) The radius of curvature of concentric shells inside/outside a spherical bubble at constant time decreases away from the bubble wall on both sides. In this case, observers on both sides of the bubble believe that they are actually on the inside of an expanding bubble, and the $\Lambda_f = 0$ bubble never grows to encompass the observer in the $\Lambda_i > 0$ region. As a result, the observer in the $\Lambda_i > 0$ region never enters the past horizon of the $\Lambda_f = 0$ census taker, so he will never be counted in her census.

The question of whether the observer in the $\Lambda_i > 0$ region sees the domain wall approaching or receding depends on the relative sizes of $\Lambda_i$ and the tension $T$ of the domain wall. In particular, the qualification requirement for the census is given in 4d reduced Planck units by \cite{Sato}
\begin{equation}
T \leq \frac{2}{\sqrt{3}}\left( \sqrt{\Lambda_i} - \sqrt{\Lambda_f }  \right)\,,
\label{WGCdomain}
\end{equation}
where in the case at hand, $\Lambda_f = 0$.
If this bound is satisfied, the $\Lambda_i > 0$ observer will eventually enter the past light cone of the maximal census taker. If it is violated, the $\Lambda_i > 0$ observer will remain forever beyond her horizon, and information about the $\Lambda_i > 0$ observer can reach the maximal census taker only in a highly scrambled form through the radiation she receives from her horizon \cite{Bousso:2011up}.

This bound is familiar from the classic paper on vacuum decay by Coleman and de Luccia \cite{Coleman:1980aw}: when \eqref{WGCdomain} is satisfied comfortably, gravitational effects are weak and can be largely neglected from the computation of the decay rate. When it is violated, gravitational effects are strong. Indeed, for a decay from a $\Lambda_i = 0$ vacuum to a $\Lambda_f < 0$ vacuum, these gravitational effects become so strong that bubble nucleation shuts off entirely. BPS domain walls in the supersymmetric $\Lambda_i = 0$ vacuum saturate this bound \cite{Cvetic:1993xe}, and their decay rate vanishes, so the supersymmetric vacuum is (marginally) stable.

An upper bound on the tension of a brane which is saturated by a BPS brane smells an awful lot like a Weak Gravity Conjecture, and indeed this connection was discussed at length by Freivogel and Kleban in \cite{Freivogel:2016qwc}, who conjectured that for every vacuum in the Landscape with vacuum energy $\Lambda_i$, there should exist another vacuum of energy $\Lambda_f < \Lambda_i$ and a domain wall between them whose tension satisfies \eqref{WGCdomain}. If true, this conjecture would imply that the lifetime of any de Sitter vacuum must be shorter than its Poincar\'e recurrence time, and it would imply that any non-supersymmetric Anti-de Sitter vacuum will decay in finite time (see also \cite{Ooguri:2016pdq}).

The relationship between this conjecture and the usual Weak Gravity Conjecture for $p$-form gauge fields can be made more concrete in the case of domain walls separating vacua distinguished by different values of flux \cite{Freivogel:2016qwc, Ooguri:2016pdq}. In particular, consider the 4d description of axion monodromy inflation \cite{Mcallister:2008hb, Silverstein:2008sg} pioneered in \cite{Kaloper:2008fb, Kaloper:2011jz}, which has a Lagrangian of the form
\begin{equation}
\mathcal{L} = \frac{1}{2} (\partial_\mu \phi)^2 - \frac{1}{2} |F_4|^2 + g \phi F_4\,,
\end{equation}
with $F_4  = d C_3$. The 3-form field has no propagating degrees of freedom in 4d, so it can be integrated out to produce a quadratic, multi-branched potential for the axion $\phi$,
\begin{equation}
V = \frac{1}{2} (n f_0 + g \phi)^2\,.
\end{equation}
Here, $n \in \mathbb{Z}$, and $f_0$ is the coupling constant of $C_3$. The potential is preserved under shifts $\phi \rightarrow \phi + 2 \pi f$, $n \rightarrow n - 1$ after imposing the consistency condition $2 \pi f g = f_0$.
The difference of vacuum energies between a pair of neighboring flux vacua is given roughly by $n f_0^2$, so for $n$ of order 1, \eqref{WGCdomain} becomes
\begin{equation}
T \lesssim f_0 \,.
\label{eq:WGC4form}
\end{equation}
Up to an unspecified $O(1)$ prefactor, this is simply the WGC bound for a domain wall charged under a 3-form gauge field. 

Why is this significant? In past discussions, the fact that the observations of a census taker in a hat region could depend on ``arcane details'' such as the tension of a particular domain wall was viewed as an unwelcome surprise, and potentially even as an indication that such a census taker is not well-suited to making predictions in eternal inflation \cite{Freivogel:2011eg}.
However, recent work on the WGC suggests a very different perspective: we have seen in many contexts that the violation/observation of a WGC bound is \emph{not} an arcane detail, but rather a crucial consistency condition of quantum gravity. In a theory which violates the ordinary WGC bound $q/m \geq \gamma_d$, non-supersymmetric extremal black holes cannot decay. Analogously, we see here that a $\Lambda_i > 0$ vacuum decay to a $\Lambda_f = 0$ vacuum cannot occur in the past light cone of a census taker in the latter vacuum if no domain wall between them satisfies the WGC-like bound \eqref{WGCdomain}.

Perhaps, like the ordinary WGC, this should be interpreted as a consistency condition on the effective field theory of the $\Lambda_i > 0$ vacuum: given a $\Lambda_i > 0$ vacuum, there necessarily exists a $\Lambda_f = 0$ vacuum and a domain wall separating the two vacua whose tension satisfies \eqref{WGCdomain}. Such a condition would represent a strengthening of the conjecture of \cite{Freivogel:2016qwc}, which demands the existence of a domain wall with tension satisfying \eqref{WGCdomain} for every $\Lambda_i > 0$, but does not demand $\Lambda_f= 0$. This is also related to the conjecture of McNamara and Vafa that there should be no nontrivial cobordisms in quantum gravity \cite{McNamara:2019rup}, which implies that there must exist a domain wall between any two vacua but does not impose a constraint on the tension of this domain wall.

On the other hand, it is quite hard to imagine that such a decay channel should exist for any $\Lambda_i > 0$ vacuum: it seems very unlikely, for instance, that our standard model vacuum can decay to a $\Lambda_f = 0$ vacuum via a domain wall whose tension satisfies \eqref{WGCdomain}, though it is difficult to rule out this possibility without a better understanding of the quantum gravity Landscape. In addition, it is worth noting that that the condition \eqref{WGCdomain} is not satisfied by the domain wall interpolating between the $\Lambda_i >0$ vacuum and the asymptotic region of scalar field space in the KKLT proposal \cite{Kachru:2003aw} (though the Strong dSC is satisfied in the asymptotic regime of scalar field space in that example). Whether this is due to an inconsistency in the KKLT proposal, indicates the existence of some domain wall not considered in the original KKLT paper, or represents a counterexample to the supposal of the preceding paragraph remains to be seen.


\section{Discussion}\label{sec:DISC}

In this paper, we have established a connection between various Swampland conjectures and the difficulty of defining asymptotic observables in expanding and eternally inflating cosmologies. We have seen that a strong version of the dSC in asymptotic regions of scalar field space---previously distinguished by dimensional reduction---would offer the best prospects for defining asymptotic observables in eternal inflation, as it implies the existence of hat regions (i.e., decelerating bubble universes) and rules out the possibility of terminal quintessence vacua in asymptotic regions of scalar field space. We have seen that the observations of a census taker in a hat region depends on whether or not a WGC-like bound is satisfied, and we have considered the possibility that this may be a consistency condition on dS vacuum decay in quantum gravity, similar to how the ordinary WGC is a consistency condition on black hole decay. We also saw, however, that a recently-proposed, sharpened version of the SDC implies that a tower of light states are thermally produced at late times in Q-space. This may be related to the definition of precise asymptotic observables in such spacetimes, which necessarily depend on thermal fluctuations, since these light massive particles may perhaps be better candidates than massless modes for the construction of an arbitrarily precise measuring device.

On the other hand, we have not tackled the most important questions: do asymptotic observables exist in expanding spacetimes? If so, what are they, and how are they computed? Can census takers in an eternally inflating universe be used to define a measure on the string Landscape, and if so, what is it? It is very possible that these questions will ultimately lead us in directions orthogonal to the the ones we have pursued in this paper: as discussed, even a decelerating cosmology suffers from asymptotic warmness, and in light of this it is not clear that even the Strong dSC is sufficient to guarantee the existence of well-defined observables. Another possibility is that our semiclassical description of expanding cosmologies must be modified significantly in the full quantum gravity, so that distant regions of space contain little new information, and asymptotic coldness is restored. This possibility is worth exploring further, especially in light of the recent realization of black hole complementarity in terms of islands \cite{Penington:2019npb, Almheiri:2019psf, Almheiri:2019hni}. Ultimately, our current cartoon picture of de Sitter and/or eternal inflation may be subject to serious revision.

Clearly, the correct picture of quantum gravity in cosmology is still coming into focus, and the ground rules are not yet clear. Nonetheless, it is heartening to think that progress in the Swampland program may shed light on the problem of asymptotic observables in cosmology, and vice versa. Certainly, these are beautiful times to ponder the mysteries of quantum gravity.

\vspace*{1cm}
{\bf Acknowledgments.}  We thank Lars Aalsma, David Andriot, Raphael Bousso, Juan Carlos Carrasco Mart\'inez, Hugo Marrochio, Liam McAllister, Jacob McNamara, Miguel Montero, Matthew Reece, Gary Shiu, Eva Silverstein, Cumrun Vafa, Irene Valenzuela, and Thomas Van Riet for useful discussions. We further thank Lars Aalsma, Raphael Bousso, Hugo Marrochio, Liam McAllister, Matthew Reece, and Gary Shiu for comments on a draft. We thank the String Phenomenology 2021 conference, where some of these useful discussions took place. The work of TR was supported by NSF grant  PHY1820912, the Simons Foundation, and the Berkeley Center for Theoretical Physics.

\bibliography{ref}

\end{document}